\documentclass[twocolumn,preprintnumbers,amsmath,amssymb,aps,prb]{revtex4}
\usepackage{amsfonts}
\usepackage{amsmath}
\usepackage{amssymb}
\usepackage{graphicx}%
\providecommand{\U}[1]{\protect\rule{.1in}{.1in}}

\begin{document}
\preprint{ }
\title{
Finite-temperature Wigner solid and other phases of ripplonic polarons \\ 
on a helium film
}

\author{
S.~N.~Klimin$^{1}$, 
J.~Tempere$^{1,2}$,
V.~R.~Misko$^{1,3}$, and
M.~Wouters$^{1}$
}
\affiliation{
$^{1}$TQC, Universiteit Antwerpen, Universiteitsplein 1, B-2610 Antwerpen, Belgium \\
$^{2}$Lyman Laboratory of Physics, Harvard University, USA \\
$^{3}$Departement Fysica, Universiteit Antwerpen, Groenenborgerlaan 171, B-2020 Antwerpen, Belgium
}

\begin{abstract}
Electrons on liquid helium can form different phases depending on density,
and temperature. Also the electron-ripplon coupling strength influences
the phase diagram, through the formation of so-called ``ripplonic
polarons'', that change how electrons are localized, and that shifts the
transition between the Wigner solid and the liquid phase. We use an
all-coupling, finite-temperature variational method to study the formation
of a ripplopolaron Wigner solid on a liquid helium film for different
regimes of the electron-ripplon coupling strength. In addition to the
three known phases of the ripplopolaron system (electron Wigner solid, polaron
Wigner solid, and electron fluid), we define and identify a fourth
distinct phase, the ripplopolaron liquid. We analyse the transitions
between these four phases and calculate the corresponding phase diagrams.
This reveals a reentrant melting of the electron solid as a function of
temperature. The calculated regions of existence of the Wigner solid are
in agreement with recent experimental data.
\end{abstract}
\date{\today}
\maketitle

\section{Introduction}

The two-dimensional (2D) electron system formed on the surface of liquid
helium has been widely investigated, especially with regard to the formation
and melting of a Wigner crystal, or Wigner solid (WS)~\cite{AndreiBook}. In
the WS phase, the electrons are self-trapped in a commensurate surface
deformation of liquid $^{4}$He called the dimple lattice
\cite{monarkha1975,fisher1979,saitoh1986}. The self-trapping effect of the
surface electrons is similar to the formation of polaron states where
electrons are dressed by self-induced lattice deformations, or virtual
phonons~\cite{jtd2010,dykman2015}.

Being driven by a force parallel to the surface of liquid $^{4}$He, the WS
moves as a whole keeping the hexagonal order. The electron motion on liquid
helium is associated with surface excitations, or ripplons (see,
e.g.,~\cite{Tempere}). When traveling faster than the ripplon phase velocity,
as in the case of the Cherenkov radiation, an electron radiates surface waves
and the ripplons emitted by different electrons interfere constructively if
the wave number of the ripplons equals the reciprocal lattice vector of the
Wigner solid (the Bragg condition). This resonant Bragg-Cherenkov emission of
ripplons gives rise to the limitation of the electron
velocity~\cite{Kristensen,Dykman}. Another intriguing nonlinear phenomenon is
a sharp rise in mobility at a much higher velocity~\cite{Shirahama,Ikegami}
which was attributed to the \textit{decoupling} of the WS from the dimple
lattice. This decoupling can be explained within a hydrodynamic
model~\cite{Vinen} assuming that the dimple lattice deepens due to the
Bragg-Cherenkov scattering which bridges the two above-mentioned phenomena.

One of the most actively developing research directions in the field, which
became possible due to the recent advances in the microfabrication technology,
is the study of the Wigner solid in confined geometries using devices such as
microchannel arrays~\cite{Glasson,Ikegami}, single-electron
traps~\cite{Papageorgiou}, field-effect transistors (FET)~\cite{Klier} and
charge-coupled devices~\cite{Sabouret}. One of the advances in this direction
was accessing the \textquotedblleft quantum wire\textquotedblright\ regime,
when the effective width of a conductive channel is less than the thermal
wavelength of the electrons. This was achieved in the recent
experiments~\cite{Rees-JLTP-2010,Rees-PRL-2011,Rees-PRL-2012,Rees-JPSJ} where
the transport properties of electrons were measured in a microchannel with the
confinement potential controlled on the scale of the inter-electron separation
($\approx0.5%
\operatorname{\mu m}%
$). {\normalsize Note that in these experimental studies, electrons are
confined in channels with constrictions. In our theoretical model developed
for an infinite system we only used typical experimental values of the
electron density, }$n_{0}${\normalsize , and the thickness of the helium film
(helium depth), }$h${\normalsize .}

The motion of electrons, or in general charged particles, in
quasi-one-dimensional (Q1D) channels has been analyzed, using numerical
simulations, in early works. For example, the structural, dynamic properties
and melting of a Q1D system of charged particles, interacting through a
screened Coulomb potential were studied~\cite{Piacente2004,Piacente2005} using
Monte Carlo simulations. However, the
experiments~\cite{Rees-JLTP-2010,Rees-PRL-2011,Rees-PRL-2012} revealed new
interesting behavior, such as oscillations in the single-electron conductance
in short and long constrictions, which required understanding and therefore
stimulated new theoretical and numerical studies. Thus, step-like electric
conduction of a classical 2D electron system in a microchannel with a narrow
constriction has been analyzed~\cite{Araki-Hayakawa-2012}. Related numerical
studies~\cite{anna-BRL,anna-EPJB}, using molecular-dynamics simulations of
Langevin equations of motion of interacting electrons on surface of liquid
$^{4}$He, revealed a significant difference in the electron dynamics for long
and short constrictions. The pronounced current oscillations found for a short
constriction were shown to be suppressed for longer
constrictions~\cite{anna-EPJB,anna-BRL}, in agreement with the experimental
observations. Also, an asymmetric FET-like structure has been
proposed~\cite{anna-EPJB} that allows an easy control of relatively large
electron flows and can be used for rectification of an ac-driven electron
flow. Furthermore, the authors~\cite{anna-EPJB,anna-BRL} addressed the
important issue of the so-called \textquotedblleft non-sequential ordering of
transitions (non-SOT)\textquotedblright\ characterized by inversions in the
subsequent number of electron rows in a Q1D channel, e.g., \textquotedblleft%
1-2-4-3\textquotedblright\ (see, e.g.,~\cite{Piacente2005}). In particular,
they found the sequence of transitions \textquotedblleft%
1-2-4-3-6-4-5\textquotedblright\ with two striking inversions
\textquotedblleft2-4-3\textquotedblright\ and \textquotedblleft%
3-6-4\textquotedblright~\cite{anna-EPJB} and demonstrated that some amount of
fluctuations (i.e., in the number of particles) restores the usual sequential
order, i.e., \textquotedblleft1-2-3-4-5\textquotedblright. The role of the
potential profile and the form of the interparticle interaction (e.g., the
screening length for electrons) in the appearance of the non-SOT has been
recently further analyzed in detail~\cite{SOT-PRB-2014,SOT-PRB-2015}.

Despite the above technological, experimental and theoretical advances in the
study of the Wigner solid, some of the fundamental properties of this system
still remain not well-understood. Moreover, recent experimental studies
revealed a number of related issues to be addressed. For example, the
experiment~\cite{Rees-PRL-2012} revealed a very gradual increase in the
electron effective mass as the temperature drops below the WS transition
temperature, while the theory predicted~\cite{Vinen} a full formation of
dimples at the transition temperature and a very weak temperature dependence.
Also, the mechanisms of the decoupling of the Wigner solid from the dimple
lattice towards an electron Wigner solid are not yet understood in detail.
{\normalsize Also, despite great efforts to observe a bound single-polaron
state experimentally, this is still an open problem. The polaronic Wigner
crystal is a well-established phenomenon, but the situation for single
polarons (which has a particular interest in view of the polaron liquid
discussed below) is not yet clear. The work on the observation of a single
polaron \cite{Andrei1984} was strongly debated.}

Here, we analyze in detail various phases of the electron-ripplon system,
i.e., when the WS is coupled to the dimple lattice, when the WS still exists
but is decoupled from the dimple lattice, and when the WS
melts, depending on such parameters of the system as the strength of the
electron-ripplon interaction, temperature and the electron concentration. Also
a polaron liquid phase is predicted in the present work at sufficiently high
temperatures combined with high coupling strengths. To the best of our
knowledge, this phase was not yet considered in the literature. The treatment
is performed within the variational scheme similar to that used in
Ref.~\cite{EPJB2003} for multielectron bubbles in liquid helium.

\section{Electron-ripplon interaction}

The Hamiltonian of a single electron on a flat helium surface is given by
\begin{align}
\hat{H} &  =\frac{\mathbf{\hat{p}}^{2}}{2m}+\sum_{q}\omega_{\mathbf{q}}\left(
\hat{a}_{\mathbf{q}}^{+}\hat{a}_{\mathbf{q}}+\frac{1}{2}\right)  \nonumber\\
&  +\frac{1}{\sqrt{S}}\sum_{\mathbf{q}}V_{\mathbf{q}}\left(  \hat
{a}_{\mathbf{q}}+\hat{a}_{-\mathbf{q}}^{+}\right)  e^{i\mathbf{q.r}%
},\label{H0}%
\end{align}
where $\mathbf{\hat{p}}$ is the electron momentum operator, $m$ is the
electron mass $S$ is the surface area, $\omega_{\mathbf{q}}$ is given by
\cite{JacksonPRB24},
\begin{equation}
\omega_{\mathbf{q}}=\sqrt{\left(  g^{\prime}q+\frac{\sigma}{\rho}q^{3}\right)
\tanh\left(  qh\right)  },\label{wq}%
\end{equation}
where $\sigma\approx3.6\times10^{-4}$ $%
\operatorname{J}%
\operatorname{m}%
^{-2}$ is the surface tension of helium, $\rho=145$ $%
\operatorname{kg}%
\operatorname{m}%
^{-3}$ is the mass density of helium, {\normalsize and }$g^{\prime}=g\left(
1+3c/\rho gh^{4}\right)  $ {\normalsize is the acceleration of the liquid due
to its van der Waals coupling to the substrate \cite{Studart} (where }$g$
{\normalsize is the acceleration due to gravity and }$c$ {\normalsize is the
van der Waals coupling of the helium to the substrate).} In the Hamiltonian
(\ref{H0}), we restrict ourselves to 2D position and momentum operators,
assuming that the part of the wave function of the electrons relating to the
direction perpendicular to the surface can be factored out exactly. The
second-quantization operators $\hat{a}_{\mathbf{q}}^{+},\hat{a}_{\mathbf{q}}$
create/annihilate a ripplon with planar wave number $\mathbf{q}$. The
electron-ripplon coupling amplitude is given by
\begin{equation}
V_{\mathbf{q}}=\sqrt{\frac{\hbar q}{2\rho\omega_{\mathbf{q}}}\tanh\left(
qh\right)  }eE,
\end{equation}
where $E$ is the electric field perpendicular to the surface (the so-called
`pressing field'), $e$ is the electron charge, and $h$ is the thickness of the
helium film. The pressing field pushes the electrons with a force $eE$ towards
the helium surface. A 1 eV barrier prevents electrons from penetrating the
helium surface. The total electric field is a sum of an external (manually
applied) field $E_{ext}$ and the electric field induced by the image charge in
the substrate with the dielectric constant $\varepsilon$:%
\begin{equation}
E=\frac{e^{2}}{4h^{2}}\frac{\varepsilon-1}{\varepsilon+1}+E_{ext}.\label{E}%
\end{equation}

It should be noted that the areas of parameters for different phases of a
ripplonic polaron system must be determined with a special care on the
stability of the system itself. For example, at very high densities, there can
exist an instability of the polaron when the pressing field becomes too large
\cite{Monarkha2012}. Also there is a maximum surface density of electrons when
a uniform distribution is stable on a flat surface \cite{GorkovJETP18}.

The self-induced trapping potential of the electron on the helium surface is
manifested by the appearance of a dimple in the helium surface underneath the
electron, much like the deformation of a rubber sheet when a person is pulled
down on it by a gravitational force. The resulting quasiparticle consists of
the electron together with its dimple and can be called a ripplonic polaron or
ripplopolaron \cite{ShikinJETP38}.

The Hamiltonian (\ref{H0}) for the ripplopolarons is very similar to the
Fr\"{o}hlich Hamiltonian describing polarons \cite{FrohlichAdvP3}; the role of
the phonons is now played by the ripplons. Methods suitable for the study of
single polarons have been used to analyze the single polaron on a flat surface
\cite{JacksonPRB24,JacksonPRB30}. The path integral treatment for a Wigner
solid of polarons has been developed in Refs. \cite{FratiniEPJB14,Rastelli}.
In Ref. \cite{EPJB2003}, we adapt their method so that it becomes suitable for
the treatment of a lattice of ripplopolarons in multielectron bubbles.

\section{Hamiltonian for a ripplopolaron in a Wigner solid}

\label{sec:1}

In their treatment of the electron Wigner solid embedded in a polarizable
medium such as a semiconductors or an ionic solid, Fratini and Qu\'{e}merais
\cite{FratiniEPJB14} described the effect of the electrons on a particular
electron through a mean-field lattice potential. The (classical) lattice
potential $V_{lat}$ is obtained by approximating all the electrons acting on
one particular electron by a homogenous charge density in which a hole is
punched out; this hole is centered in the lattice point of the particular
electron under investigation and has a radius given by the lattice distance
$d$. Thus, in their approach, anisotropy effects, e.~g., related to the
lattice orientation, are neglected. A second assumption implicit in this
approach is that the effects of exchange are neglected. This can be justified
by noting that for the electrons to form a Wigner solid it is required that
their wave function is localized to within a fraction of the lattice parameter
as follows from the Lindemann criterion \cite{LindemanZPhys11}.

Within this particular mean-field approximation, the lattice potential can be
calculated from classical electrostatics and we find that for a 2D electron
gas it can be expressed in terms of the elliptic functions of first and second
kind, $E\left(  x\right)  $ and $K\left(  x\right)  $,
\begin{align}
V_{lat}\left(  \mathbf{r}\right)   &  =-\frac{2e^{2}}{\pi d^{2}}\left\{
\left\vert d-r\right\vert E\left[  -\frac{4rd}{\left(  d-r\right)  ^{2}%
}\right]  \right. \nonumber\\
&  \left.  +\left(  d+r\right)  \operatorname{sgn}\left(  d-r\right)  K\left[
-\frac{4rd}{\left(  d-r\right)  ^{2}}\right]  \right\}  . \label{Potential}%
\end{align}
Here, $\mathbf{r}$ is the position vector measured from the lattice position.
We can expand this potential around the origin to find the small-amplitude
oscillation frequency of the electron lattice:
\begin{equation}
\lim_{r\ll d}V_{lat}\left(  \mathbf{r}\right)  =-\frac{2e^{2}}{d}+\frac{1}%
{2}m\omega_{lat}^{2}r^{2}+\mathcal{O}\left(  r^{4}\right)  , \label{Potlimit}%
\end{equation}
with the confinement frequency
\begin{equation}
\omega_{lat}=\sqrt{\frac{e^{2}}{md^{3}}}. \label{phonfreq}%
\end{equation}
The `phonon' frequency $\omega_{lat}$ of the electron Wigner solid corresponds
closely to the longitudinal plasmon frequency that can be derived using an
entirely different approach based on a more rigorous study of the modes of
oscillations of both the helium surface and the charge distribution on the
surface. From this, and from the successful application of this mean-field
approach to polaron crystals in solids, we conclude that the approach based on
that of Fratini and Qu\'{e}merais describes the influence of the other
electrons well in the framework of small amplitude oscillations of the
electrons around their lattice point. The phenomenological Lindemann melting
criterion \cite{LindemanZPhys11} suggests that the Wigner solid will melt when
the electrons are on average displaced more than a certain value $\delta
_{0}<1$ from their lattice position; thus in the regime of interest the
Fratini-Qu\'{e}merais approach is applicable. In the mean-field approximation,
the Hamiltonian for a ripplopolaron in a lattice on a helium surface is given
by
\begin{align}
\hat{H}  &  ={{\frac{\hat{p}^{2}}{2m}}}+V_{lat}\left(  \mathbf{\hat{r}%
}\right)  +\sum_{\mathbf{q}}\hbar\omega_{\mathbf{q}}\hat{a}_{\mathbf{q}}%
^{+}\hat{a}_{\mathbf{q}}\nonumber\\
&  +\sum_{\mathbf{q}}V_{\mathbf{q}}e^{-i\mathbf{q.r}}\left(  \hat
{a}_{\mathbf{q}}+\hat{a}_{-\mathbf{q}}^{+}\right)  . \label{H1}%
\end{align}

\section{The ripplopolaron Wigner solid at finite temperature}

\label{sec:3}

The simple but intuitive approach of the previous section describes the system
in the limit of zero temperature. To study the ripplopolaron Wigner solid at
finite temperature (and for any value of the electron-ripplon coupling), we
use the variational path-integral approach \cite{Feynman}. This variational
principle distinguishes itself from Rayleigh-Ritz variation in that it uses a
trial action functional $S_{trial}$ instead of a trial wave function.

The action functional of the system described by Hamiltonian (\ref{H1}),
becomes, after elimination of the ripplon degrees of freedom,
\begin{align}
S &  =-{{\frac{1}{\hbar}}}\int\limits_{0}^{\hbar\beta}d\tau\left\{  {{\frac
{m}{2}}}\dot{r}^{2}(\tau)+V_{lat}[r(\tau)]\right\}  \nonumber\\
&  +\sum_{\mathbf{q}}\left\vert V_{q}\right\vert ^{2}\int\limits_{0}%
^{\hbar\beta}d\tau\int\limits_{0}^{\hbar\beta}d\sigma G_{\omega(q)}%
(\tau-\sigma)e^{i\mathbf{q}\cdot\lbrack\mathbf{r}(\tau)-\mathbf{r}(\sigma
)]},\label{S}%
\end{align}
with
\begin{equation}
G_{\nu}(\tau-\sigma)={{\frac{\cosh[\nu(|\tau-\sigma|-\hbar\beta/2)]}%
{\sinh(\beta\hbar\nu/2)}}}.
\end{equation}
In preparation of its customary use in the Jensen-Feynman inequality, the
action functional (\ref{S}) is written in imaginary time $t=i\tau$ with
$\beta=1/(k_{B}T)$ where $T$\thinspace is the temperature. Following an
approach analogous for a lattice of polarons in an ionic crystal
\cite{FratiniEPJB14,Rastelli}, and to that of Devreese et al. for $N$ polarons
in a quantum dot \cite{DevreeseSSC114}, we introduce a quadratic trial action
of the form
\begin{align}
S_{trial} &  =-{{\frac{1}{\hbar}}}\int\limits_{0}^{\hbar\beta}d\tau\left[
{{\frac{m}{2}}}\dot{r}^{2}(\tau)+{{\frac{m\Omega^{2}}{2}}}r^{2}(\tau)\right]
\nonumber\\
&  -{{\frac{Mw^{2}}{4\hbar}}}\int\limits_{0}^{\hbar\beta}d\tau\int
\limits_{0}^{\hbar\beta}d\sigma G_{w}(\tau-\sigma)\mathbf{r}(\tau
)\cdot\mathbf{r}(\sigma).\label{S0}%
\end{align}
where $M,w,$ and $\Omega$ are the variationally adjustable parameters. This
trial action corresponds to the Lagrangian
\begin{equation}
\mathcal{L}_{0}={{\frac{m}{2}}}\dot{r}^{2}+{{\frac{M}{2}}}\dot{R}^{2}%
-{{\frac{\kappa}{2}}}r^{2}-{{\frac{K}{2}}}(\mathbf{r}-\mathbf{R}%
)^{2},\label{L0}%
\end{equation}
from which the degrees of freedom associated with $\mathbf{R}$ have been
integrated out. This Lagrangian can be interpreted as describing an electron
with mass $m$ at position $\mathbf{r}$, coupled through a spring with spring
constant $\kappa$ to its lattice site, and to which a fictitious mass $M$ at
position $\mathbf{R}$ has been attached with another spring, with spring
constant $K$. The relation between the spring constants in (\ref{L0}) and the
variational parameters $w,\Omega$ is given by
\begin{align}
w &  =\sqrt{K/m},\\
\Omega &  =\sqrt{(\kappa+K)/m}.
\end{align}

Based on the trial action $S_{trial}$, the Jensen-Feynman variational method
allows one to obtain an upper bound for the free energy $F$ of the system (at
temperature $T$) described by the action functional $S$ by minimizing the
following function:%

\begin{equation}
F_{var}=F_{0}-\frac{1}{\beta}\left\langle S-S_{trial}\right\rangle
_{S_{trial}}, \label{JF}%
\end{equation}
with respect to the variational parameters of the trial action. In this
expression, $F_{0}$ is the free energy of the trial system characterized by
the action $S_{trial}$, $\beta=1/(k_{b}T)$ is the inverse temperature, and the
expectation value $\left\langle S-S_{trial}\right\rangle _{S_{trial}}$ is to
be taken with respect to the ground state of this trial system.

The evaluation of expression (\ref{JF}) is straightforward though lengthy. We
find%
\begin{align}
\lefteqn{F_{var}=\frac{2}{\beta}\ln\left[  2\sinh\left(  \frac{\beta
\hbar\Omega_{1}}{2}\right)  \right]  +\frac{2}{\beta}\ln\left[  2\sinh\left(
\frac{\beta\hbar\Omega_{2}}{2}\right)  \right]  }\nonumber\\
&  -{{\frac{2}{\beta}}}\ln\left[  2\sinh\left(  {{\frac{\beta\hbar w}{2}}%
}\right)  \right]  -{{\frac{\hbar}{2}}}\sum_{i=1}^{2}a_{i}^{2}\Omega_{i}%
\coth\left(  {{\frac{\beta\hbar\Omega_{i}}{2}}}\right) \nonumber\\
&  -{{\frac{\sqrt{\pi}e^{2}}{\sqrt{D_{0}}}}}e^{-d^{2}/(2D_{0})}\left[
I_{0}\left(  {{\frac{d^{2}}{2D_{0}}}}\right)  +I_{1}\left(  {{\frac{d^{2}%
}{2D_{0}}}}\right)  \right] \nonumber\\
&  -{{\frac{1}{2\pi\hbar}}}\int_{0}^{k_{c}}dq~q|V_{q}|^{2}\int_{0}^{\hbar
\beta/2}d\tau{{\frac{\cosh[\omega_{q}(\tau-\hbar\beta/2)]}{\sinh[\beta
\hbar\omega_{q}/2]}}}\nonumber\\
&  \times\exp\left[  -{{\frac{q^{2}}{2}}}\left(  D_{0}-D_{\tau}\right)
\right]  . \label{F}%
\end{align}
In this expression, $I_{0}$ and $I_{1}$ are Bessel functions of imaginary
argument, $k_{c}=\left(  \rho g^{\prime}/\sigma\right)  ^{1/2}$ is the
capillary wave number \cite{JacksonPRB24}. The capillary wave number serves as
a cutoff in the integral over $q$ for the polaron free energy. The function
$D_{\tau}$ is given by:
\begin{equation}
D_{\tau}={{\frac{\hbar}{m}}}\sum_{j=1}^{2}\frac{a_{j}^{2}}{\Omega_{j}}%
{{\frac{\cosh[\hbar\Omega_{j}(\tau-\beta/2)]}{\sinh(\hbar\Omega_{j}\beta/2)}}%
},
\end{equation}
with the coefficients%
\begin{equation}
a_{1}=\sqrt{{{\frac{\Omega_{1}^{2}-w^{2}}{\Omega_{1}^{2}-\Omega_{2}^{2}}}}%
};\quad a_{2}=\sqrt{{{\frac{w^{2}-\Omega_{2}^{2}}{\Omega_{1}^{2}-\Omega
_{2}^{2}}}}}.
\end{equation}
The frequencies $\Omega_{i}$ $\left(  i=1,2\right)  $ are the eigenfrequencies
of the trial system, and $w$ is the third (auxiliary) frequency which is also
the variational parameter. The parameter $d$ is the inter-electron distance on
the helium surface, which is related to the concentration as:%
\[
d=\frac{2}{\left(  \sqrt{3}n_{0}\right)  ^{1/2}}.
\]

The two first lines in the expression (\ref{F}) for the variational free
energy describe the free energy of the model system and the averaged influence
phase of the model system. The third line in (\ref{F}) is the averaged energy
of the Coulomb interaction of the electron with the self-consistent field
induced by other electrons. In other words, this is the averaged potential
energy of the electron in the Wigner solid. Finally, the last line is the
polaron contribution to the free energy.

Optimal values of the variational parameters are determined by numerical
minimization of the variational functional $F$ as given by expression
(\ref{F}). The result of the variational path-integral method allows us to
introduce different measurable quantities, e. g., temperature to examine the
melting of the Wigner solid, and the effective mass of a polaron. The latter
one can be estimated as $\left(  m+M\right)  $ where $M$ is the mass of the
fictitious particle. A signature of the polaron phase can be a drastic change
of the mobility when varying the coupling strength, that can allow one to
distinguish between polaron and electron WS experimentally.

\section{Results and discussion}

In this section, we calculate different parameters for a ripplopolaron Wigner
solid on the liquid helium surface. This is performed using the variational
approach for the polaron free energy as described above. Optimal values of the
variational parameters are determined by the numerical minimization of the
variational functional $F_{var}$ given by expression (\ref{F}). We can
consider, as a starting point for the treatment, the experimental conditions
as obtained from Ref. \cite{Rees-PRL-2011}, where the thickness of the helium
film was $h\approx1%
\operatorname{\mu m}%
$ to $h\approx1.7%
\operatorname{\mu m}%
$, the temperature was $T\approx1%
\operatorname{K}%
$ to $T\approx1.2%
\operatorname{K}%
$, and the concentration of electrons on the surface was $n_{0}\approx
2.58\times10^{9}%
\operatorname{cm}%
^{-2}$ and $n_{0}\approx3.03\times10^{9}%
\operatorname{cm}%
^{-2}$. We however vary temperatures and concentrations in a rather wide range
around those values.

\begin{figure}
[th]
\begin{center}
\includegraphics[
height=2.1932in,
width=2.9395in
]%
{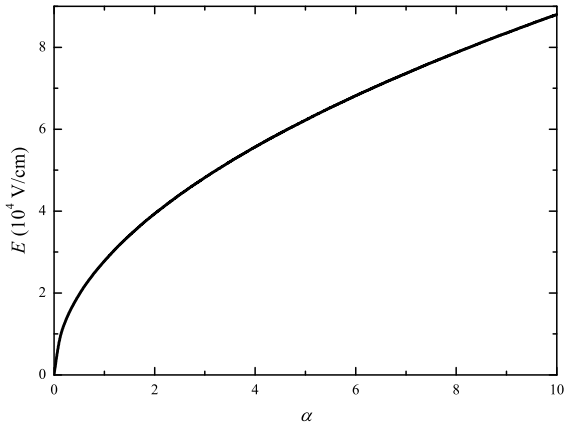}%
\caption{Dependence of the electric field $E$ measured in V/cm on the
dimensionless electron-ripplon coupling constant $\alpha$.}%
\label{fig:ealpha}%
\end{center}
\end{figure}

The electron-ripplon coupling is measured through the dimensionless coupling
constant $\alpha$ determined as \cite{fisher1979}:%
\begin{equation}
\alpha=\frac{\left(  eE\right)  ^{2}}{8\pi\sigma}\frac{2m}{\hbar^{2}k_{c}^{2}%
},\label{alpha}%
\end{equation}
where $E$ is the electric field applied perpendicular to the surface. It
includes both the image field induced by a polar substrate and an external
field which can be controlled artificially. Fig. \ref{fig:ealpha} shows the
correspondence between $E$ and $\alpha$ for the aforesaid set of material
parameters. Note that for $h\approx1%
\operatorname{\mu m}%
$, the contribution to $\alpha$ from the image field is negligibly small: even
with a metallic substrate, $\alpha\lesssim10^{-7}$, so that the
electron-ripplon coupling can be completely controlled by an external field.%

For the numeric calculation, the dimensionless units are used with $\hbar=1$,
the electron mass $m=1$ and the unit for the energy is $\frac{\hbar^{2}%
k_{c}^{2}}{2m}=1$, where $k_{c}\approx6\times10^{5}%
\operatorname{cm}%
^{-1}$ is the capillary wave number from Ref. \cite{JacksonPRB24}.
{\normalsize Also the effective acceleration }$g^{\prime}=10^{8}g$
{\normalsize is taken from Ref. \cite{JacksonPRB24}. Note that, despite a
substantial dependence of }$g^{\prime}$ {\normalsize on the helium film
thickness, this dependence can only slightly change the phase diagrams
calculated below, {\normalsize because at given }}$\alpha$,
{\normalsize {\normalsize it influences the results only through ripplon
frequencies} (which are very small for any reasonable }$g^{\prime}%
${\normalsize ).}

The polaronic aspects in the formation of the electron Wigner solid on the
liquid helium surface are already thoroughly studied both experimentally and
theoretically, see, e. g., the recent review \cite{Monarkha2012} and
references therein. However, some questions remain unexplored. The transition
between two types of the Wigner solid (the electron and polaron Wigner solid)
at different temperatures is of a particular interest, because this problem
requires an arbitrary-coupling finite-temperature polaron theory. We
successfully applied this polaron theory to investigate polaron Wigner solids
in multielectron bubbles. Here, the same approach is used for the calculation
of the phase diagrams on the flat helium surface.

The phenomenological Lindemann criterion \cite{LindemanZPhys11} is frequently
used in the literature for the determination of a melting point in a Wigner
solid. This criterion states in general that a crystal lattice of objects (be
it atoms, molecules, electrons, or polarons) will melt when the average motion
of the objects $\sqrt{\left\langle \mathbf{r}^{2}\right\rangle }$ around their
lattice site is larger than a critical fraction $\delta_{0}$ of the lattice
parameter $d$. It would be very hard to calculate from first principles the
exact value of the critical fraction $\delta_{0}$, but for the particular case
of electrons on a helium surface, we can make use of an experimental
determination. Grimes and Adams \cite{GrimesPRL42} found that the Wigner solid
melts when $\Gamma=137\pm15$, where $\Gamma$ is the ratio of potential energy
to the kinetic energy per electron. In their experiment, the electron density
varied from $10^{8}$ cm$^{-2}$ to $3\times10^{8}$ cm$^{-2}$ while the melting
temperature $T_{c}$ varied from 0.23 K to 0.66 K. As estimated in Ref.
\cite{EPJB2003} using the experimental data by Grimes and Adams
\cite{GrimesPRL42}, the critical fraction equals $\delta_{0}\approx0.13$.
Recently, a modified Lindemann criterion has been derived in Ref.
\cite{Monarkha2012}, which is based on the calculation of a two-site
correlation function for the Wigner solid, describing the correlation of
displacements for the nearest neighbors. When combined with the Monte Carlo
calculation \cite{Bedanov}, this leads to the modified value $\delta
_{0}\approx0.212$, which is used in the present work. In Ref. \cite{EPJB2003}
we used two parallel Lindemann criteria following to the scheme developed in
Ref. \cite{FratiniEPJB14}. According to this scheme, areas of stasbility for
different phases of a ripplopolaron system are determined, at least
qualitatively, by the parameters $\delta_{c}\equiv\sqrt{\left\langle
\mathbf{R}_{c}^{2}\right\rangle }/d$ and $\delta_{\rho}\equiv\sqrt
{\left\langle \boldsymbol{\rho}^{2}\right\rangle }/d$, where $\mathbf{R}_{c}$
and $\boldsymbol{\rho}$ are, respectively, the center-of-mass and the relative
coordinate for the model polaron system. The averaged squared radii are
explicitly determined using the variational parameters for the ripplopolaron
system \cite{EPJB2003},%
\begin{align}
\left\langle \mathbf{R}_{c}^{2}\right\rangle  &  ={{\frac{w^{4}}{\left(
\Omega_{1}^{2}-\Omega_{2}^{2}\right)  \left[  \Omega_{1}^{2}\Omega_{2}%
^{2}-w^{2}(\Omega_{1}^{2}+\Omega_{2}^{2})\right]  ^{2}}}}\nonumber\\
&  \times\left[  \frac{\Omega_{2}^{4}\left(  \Omega_{1}^{2}-w^{2}\right)
}{\Omega_{1}}\coth\left(  \frac{\beta\Omega_{1}}{2}\right)  \right.
,\nonumber\\
&  \left.  +\frac{\Omega_{1}^{4}\left(  w^{2}-\Omega_{2}^{2}\right)  }%
{\Omega_{2}}\coth\left(  \frac{\beta\Omega_{2}}{2}\right)  \right]
,\label{Rc}\\
\left\langle \boldsymbol{\rho}^{2}\right\rangle  &  ={{\frac{1}{\Omega_{1}%
^{2}-\Omega_{2}^{2}}}}\left[  \frac{\Omega_{1}^{3}}{\Omega_{1}^{2}-w^{2}}%
\coth\left(  \frac{\beta\Omega_{1}}{2}\right)  \right. \nonumber\\
&  \left.  +\frac{\Omega_{2}^{3}}{w^{2}-\Omega_{2}^{2}}\coth\left(
\frac{\beta\Omega_{2}}{2}\right)  \right]  . \label{rho}%
\end{align}

In principle, four combinations are possible:

(1) The case when both $\delta_{c}<\delta_{0}$ and $\delta_{\rho}<\delta_{0}$
corresponds to the polaron Wigner solid. In this case, electrons are strongly
localized in the Wigner solid together with the dimple lattice on the helium
surface. This regime is called in Ref. \cite{Monarkha2012} the polaron
anchoring of the Wigner crystal.

(2) When $\delta_{c}<\delta_{0}$ and $\delta_{\rho}>\delta_{0}$, the electron
Wigner solid exists without anchoring to dimples (electrons leave dimples but
the Wigner solid still exists). However, the parameters of this system are
still influenced by the electron-phonon interaction through scattering of
ripplons on the electrons. Therefore, this regime can be considered as the
electron Wigner solid. In other words, this is the Wigner solid of
weak-coupling polarons.

(3) When $\delta_{c}>\delta_{0}$ and $\delta_{\rho}<\delta_{0}$, the ripplonic
polarons are chaotically moving but electrons are in dimples. This case can be
interpreted as a polaron liquid. It can be realized when both the coupling
strength and the temperature are sufficiently high.

(4) When both $\delta_{c}>\delta_{0}$ and $\delta_{\rho}>\delta_{0}$, this is
the case when the Wigner solid melts to the electron liquid (with the polaron effect).%

\begin{figure}
[th]
\begin{center}
\includegraphics[
height=5.7614in,
width=2.8746in
]%
{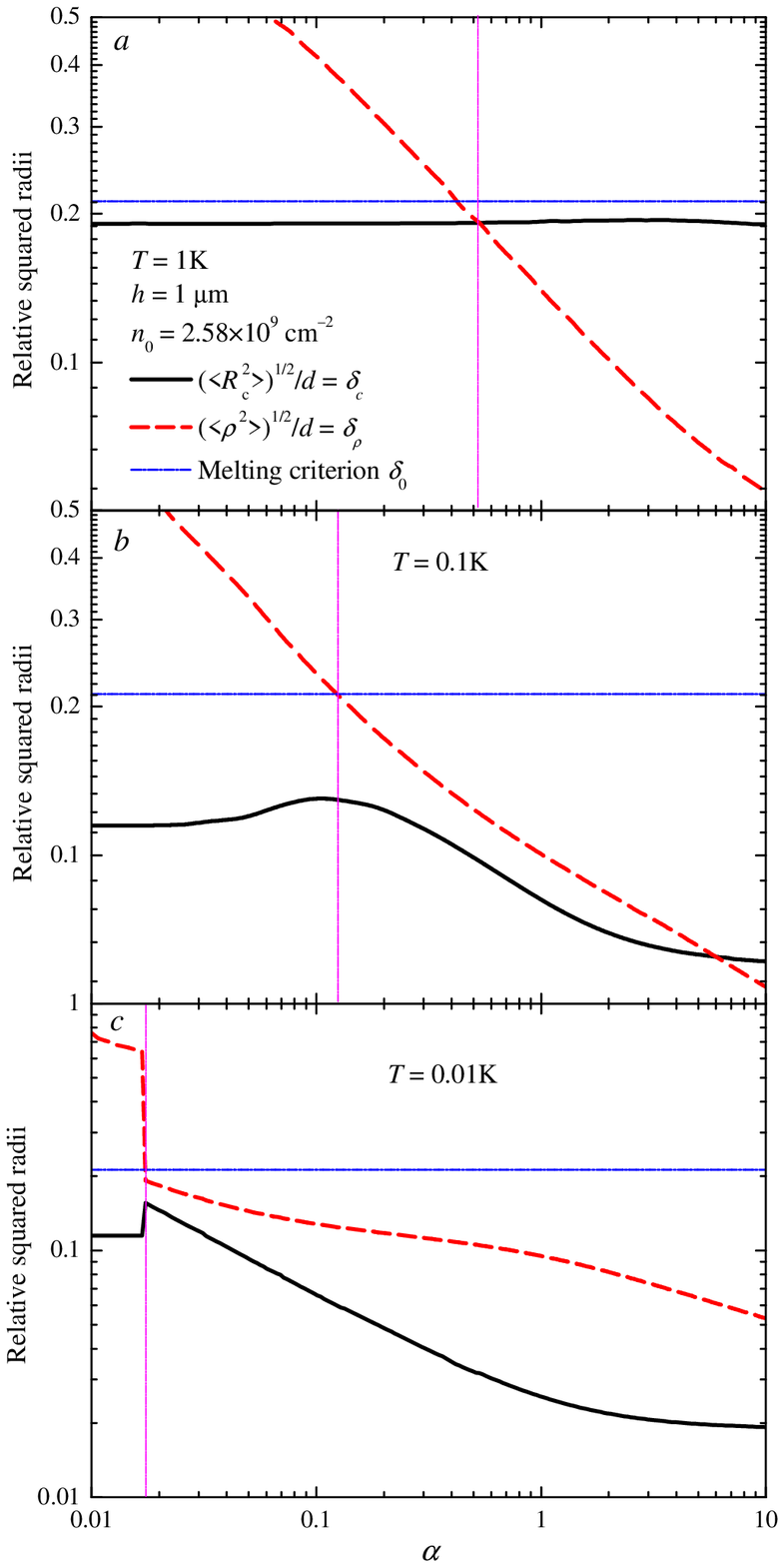}%
\caption{Parameters $\delta_{c}$ (solid curves) and $\delta_{\rho}$ (dashed
curves) for a ripplopolaron Wigner lattice as a function of the coupling
constant $\alpha$ for the electron system on the surface of the helium film of
the width $h=1\operatorname{\mu m}$. The concentration of electrons is
$n_{0}=2.58\times10^{9}\operatorname{cm}^{-2}$, the temperatures are
$T=1\operatorname{K}$ (\emph{a}), $T=0.1\operatorname{K}$ (\emph{b}), and
$T=0.01\operatorname{K}$ (\emph{c}). The dot-dashed line shows the critical
value $\delta_{0}=0.212$. The vertical lines indicate a transition between
polaron and electron Wigner lattices.}%
\label{fig:radii}%
\end{center}
\end{figure}

In Fig. \ref{fig:radii} (\emph{a}), we plot parameters $\delta_{c}$ (the solid
curve) and $\delta_{\rho}$ (the dashed curve) for a ripplopolaron Wigner solid
as a function of the coupling constant $\alpha$ for the electron system on the
surface of the helium film. The dot-dashed curve shows the critical value for
the modified Lindemann melting criterion $\delta_{0}=0.212$. The thickness of
the film is $h=1%
\operatorname{\mu m}%
$. The concentration of electrons is $n_{0}=2.58\times10^{9}%
\operatorname{cm}%
^{-2}$, the temperature is $T=1%
\operatorname{K}%
$. Under these conditions, the relative averaged squared oscillation amplitude
$\delta_{\rho}$ decreases when strengthening the electron-ripplon coupling,
passing the critical value $\delta_{c}$ at $\alpha=\alpha_{c}\approx0.53$. The
center-of-mass relative averaged squared oscillation amplitude, $\delta_{c}$,
varies extremely slightly, being smaller than the critical value at all
coupling strengths. In terms of the aforesaid four regimes, this means that
the Wigner solid exists at these conditions for all $\alpha$, changing at
$\alpha=\alpha_{c}$ from the electron Wigner solid at $\alpha<\alpha_{c}$ to
the polaron Wigner solid at $\alpha>\alpha_{c}$.

{\normalsize In the experiments \cite{Rees-PRL-2011,Rees-PRL-2012,Rees-JPSJ},
where there is no additional external field to enhance }$\alpha${\normalsize ,
electrons are attracted to the surface of liquid helium by a rather small
image charge. The coupling constant in this regime is small. For example, in
the conditions of Ref. \cite{Rees-PRL-2011}, }$\alpha\sim1.5\times10^{-3}%
${\normalsize . According to Fig. \ref{fig:radii} (\emph{a}), the
ripplopolaron Wigner lattice for small } $\alpha$ {\normalsize is stable, thus
the result in Fig. \ref{fig:radii} (\emph{a}) is in line with these
experiments.}

The graphs \ref{fig:radii} (\emph{a}) and (\emph{b}) show the analogous
dependence of the parameters $\left(  \delta_{c},\delta_{\rho}\right)  $ for
lower temperatures: $T=0.1%
\operatorname{K}%
$ and $T=0.01%
\operatorname{K}%
$, respectively. For a sufficiently low temperature $T=0.01%
\operatorname{K}%
$, we can see a sharp transition between two regimes at certain $\alpha
\equiv\alpha_{c}$, which can be qualitatively attributed to a weak and
strong-coupling polaron regimes. It was found in Ref. \cite{JacksonPRB24} that
at $T=0$, there is a crossover\ between weak-coupling and strong-coupling
polaron regimes when varying $\alpha$. This transition is not discontinuous at
non-zero temperatures, although at low temperatures it can be sharp. At
$T=0.01%
\operatorname{K}%
$, as seen from Fig. \ref{fig:radii} (\emph{c}), this transition is followed
by a change of the regime for the Wigner solid: for smaller $\alpha
\lessapprox\alpha_{c}$, $\delta_{\rho}>\delta_{0}$ and $\delta_{c}<\delta_{0}%
$, so that the Wigner solid is formed by weak-coupling polarons, and for
$\alpha\gtrapprox\alpha_{c}$, we see that both $\delta_{c}$ and $\delta_{\rho
}$ are smaller than $\delta_{0}$, that corresponds to the polaron Wigner
solid. At higher temperatures, the crossover between the regimes of electron
and polaron Wigner solids is rather smooth. We can see a manifestation of this
crossover at $T=0.1%
\operatorname{K}%
$ through a non-monotonic dependence of $\delta_{c}$ as a function of the
temperature. We can also conclude from the comparison of the behavior of the
parameters $\delta_{\rho}$ and $\delta_{c}$ at different temperatures that low
temperatures are favorable for the Wigner solid formation and for its polaron anchoring.

Note that we use the same critical value $\delta_{0}$ for the melting of the
Wigner solid and for the polaron dissociation. Only for the former one, there
are experimental \cite{GrimesPRL42} and numerical \cite{Bedanov} estimates of
the Lindemann criterion, $\delta_{0}$, and even these do not agree. However,
from Fig. \ref{fig:radii} it is clear that a different choice of $\delta_{0}$
(keeping its range of magnitude) will not change the results qualitatively.

\begin{figure}
[th]
\begin{center}
\includegraphics[
height=2.2381in,
width=2.9334in
]%
{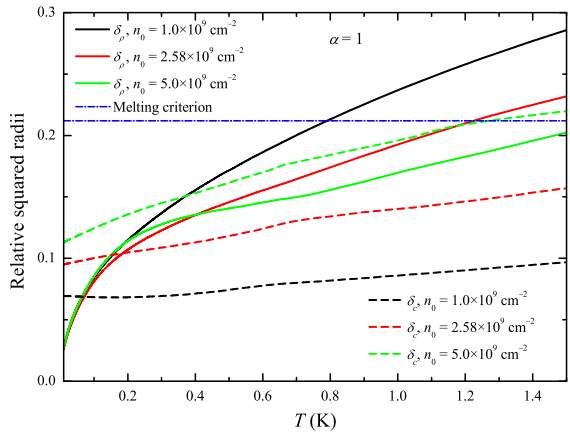}%
\caption{Parameters $\delta_{c}$ and $\delta_{\rho}$ for a ripplopolaron
Wigner lattice as a function of the temperature for the electron system on the
surface of the helium film with $\alpha=1$. the other parameters are the same
as in Fig. \ref{fig:radii}.}%
\label{fig:rad4}%
\end{center}
\end{figure}

Figure \ref{fig:rad4} shows the parameters $\delta_{c}$ and $\delta_{\rho}$
for a ripplopolaron Wigner solid at a given coupling strength $\alpha=1$ as a
function of the temperature for the ripplopolaron system on the surface of the
helium film for different electron concentrations. The other parameters are
the same as the previous figures. We can see from this figure that both
$\delta_{c}$ and $\delta_{\rho}$ increase monotonically when the temperature
rises. The parameters $\delta_{\rho}$ and $\delta_{c}$, pass the critical
Lindemann value $\delta_{0}$ at different temperatures, depending on the
electron concentration, so that the transition points between different
configurations of the ripplopolaron system depends on the concentration
$n_{0}$. For lower $n_{0}$, the transitions between different configurations
occurs at lower temperatures.%

\begin{figure}
[th]
\begin{center}
\includegraphics[
height=6.0329in,
width=3.0943in
]%
{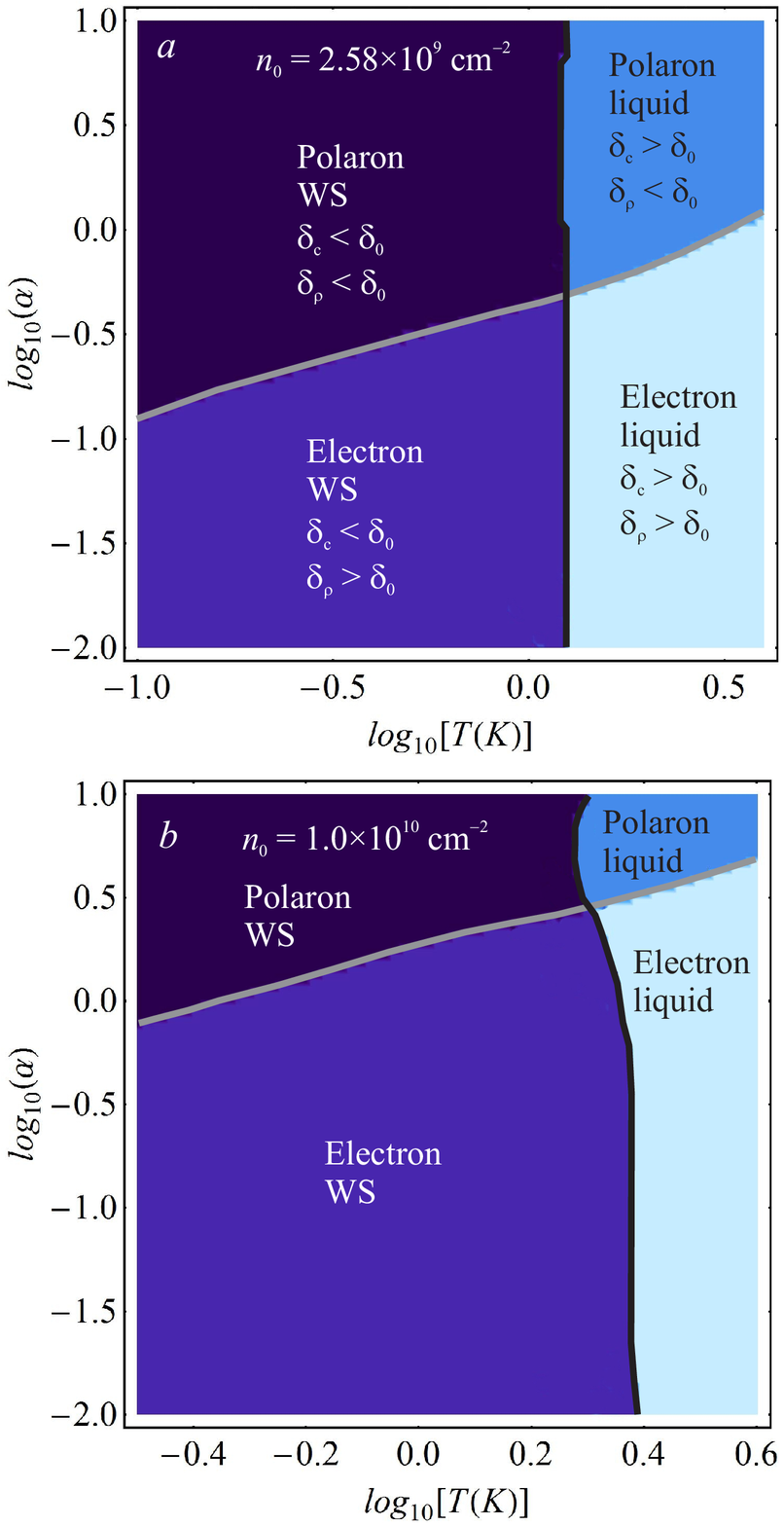}%
\caption{Phase diagram for the ripplopolaron system on the helium surface in
the variables $\left(  T,\alpha\right)  $ at a concentration of electrons
$n_{0}=2.58\times10^{9}\operatorname{cm}^{-2}$ (\emph{a}), $n_{0}%
=1.0\times10^{10}\operatorname{cm}^{-2}$ (\emph{b}), for the helium film width
$h=1\operatorname{\mu m}$.}%
\label{fig:pd1}%
\end{center}
\end{figure}

In order to obtain a more detailed picture of different regimes for
ripplopolaron Wigner solid, we calculate phase diagrams where different
regimes for the ripplopolaron system are indicated. Figure \ref{fig:pd1}
contains the phase diagram for the ripplopolaron system on the helium film
surface in the variables $\left(  T,\alpha\right)  $ (using the logarithmic
scale) calculated for two concentration of electrons $n_{0}=2.58\times10^{9}%
\operatorname{cm}%
^{-2}$ and $n_{0}=1.0\times10^{10}%
\operatorname{cm}%
^{-2}$, and for the thickness of the liquid helium film $h=1%
\operatorname{\mu m}%
$. In this figure, all four regimes described above can be seen. We can
conclude from Fig. \ref{fig:pd1} that the electron Wigner solid as obtained in
the present calculation is expected to be stable at the experimental
conditions \cite{Rees-PRL-2011}, although is rather close to the melting conditions.

It should be noted that the boundary between the regimes with $\delta
_{c}<\delta_{0}$ and $\delta_{c}>\delta_{0}$ corresponds to melting of a
Wigner crystal, i.~e., this is a true phase transition. On the contrary, the
other boundary -- between the regimes with $\delta_{\rho}<\delta_{0}$ and
$\delta_{\rho}>\delta_{0}$ indicates a transition between strong-coupling and
weak-coupling polaron states. According to the Gerlach-L\"{o}wen theorem
\cite{Gerlach}, there is no phase transition between those regimes for a
polaron. At sufficiently high temperatures $T\gtrsim0.1%
\operatorname{K}%
$ , as shown in \cite{JacksonPRB24}, the transition between strong-coupling
and weak-coupling polaron regimes is a crossover rather than a phase
transition. Correspondingly, the transition between the polaron and electron
Wigner solids is also a crossover (indicated by grey curves at the figures).

We can see from Fig. \ref{fig:pd1} that at sufficiently low densities, the
melting temperature for the transition between a polaron or electron Wigner
solid to a polaron or electron liquid only weakly depends on the
electron-ripplon coupling constant $\alpha$, and it becomes more sensitive to
$\alpha$ at higher densities. This weak coupling depencence of the melting
temperature can be explained by the fact that an overlap of polaron dimples at
low densities is relatively small, increasing when rising density. The melting
temperature is a non-monotonic function of $\alpha$, which is one of
manifestations of the reentrant melting discussed below.%

The other boundary at the phase diagrams in Fig. \ref{fig:pd1}, which
corresponds to the polaron dissociation, behaves as follows. When the coupling
strength gradually increases at a sufficiently low temperature, the regime of
the electron Wigner solid turns at a certain $\alpha$ to the polaron Wigner
solid. This critical $\alpha$ rises when increasing temperature. At higher
temperatures, when increasing $\alpha$, the electron liquid can change to the
polaron liquid without forming a polaron Wigner solid. It is often assumed
that the formation of polaron dimples always leads to their Wigner
crystallization. However, according to the present variational calculation,
there exists a regime where the polarons are not yet dissociated but their
mass is not large enough to form a Wigner crystal. As seen from Fig.
\ref{fig:pd1}, it requires a combination of large $\alpha$ and high
temperatures. This transition was predicted for an electron-phonon system in a
3D polar crystal \cite{FratiniEPJB14}. For a ripplonic polaron system, to the
best of our knowledge, this regime was not yet discussed in the literature.

\begin{figure}
[th]
\begin{center}
\includegraphics[
height=5.9759in,
width=3.1341in
]%
{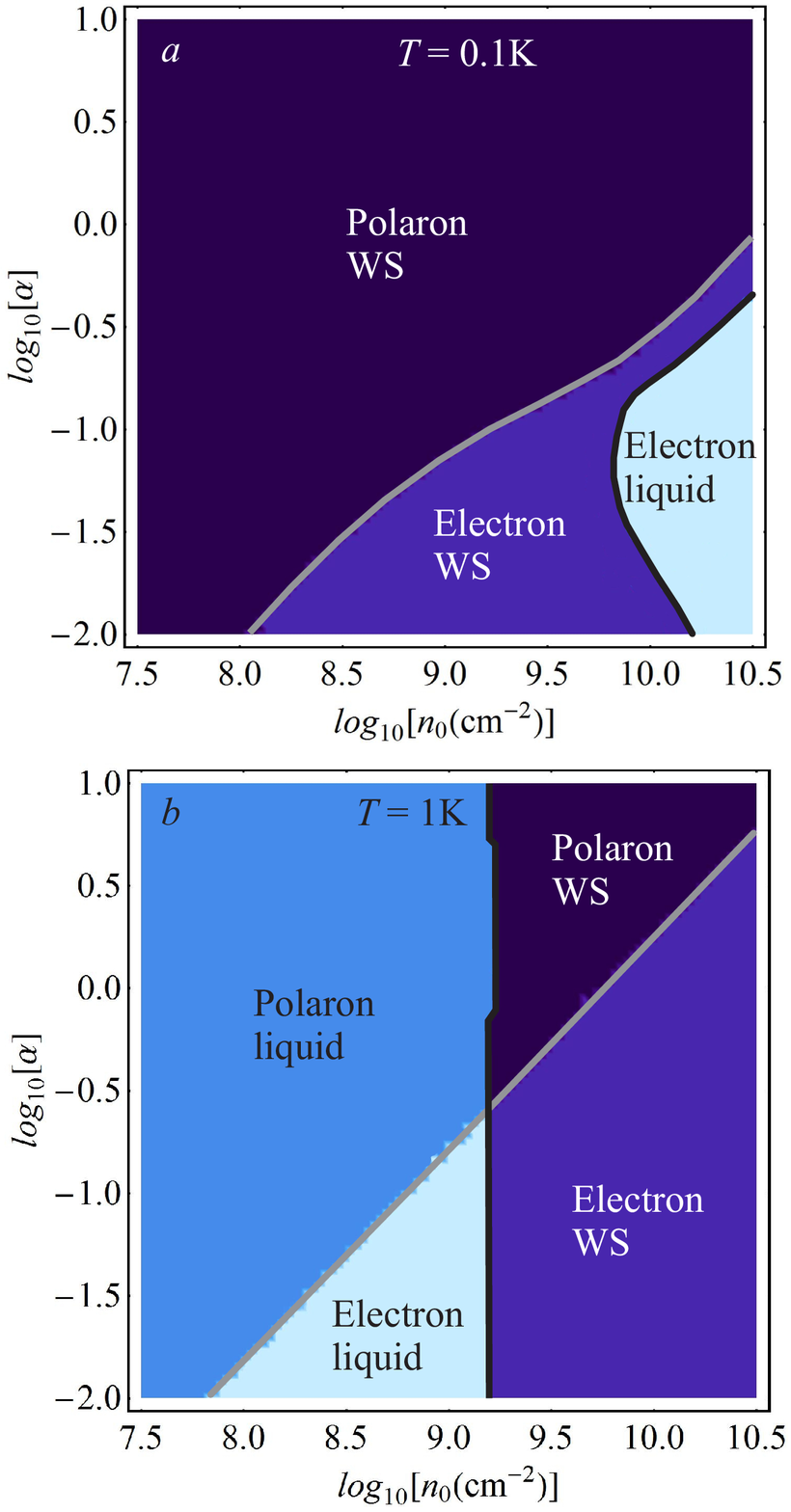}%
\caption{Phase diagram for the ripplopolaron system on the helium surface in
the variables $\left(  n_{0},\alpha\right)  $ for the temperature
$T=0.1\operatorname{K}$ (\emph{a}) and $T=1\operatorname{K}$ (\emph{b}), with
the helium film width $h=1\operatorname{\mu m}$.}%
\label{fig:pd2}%
\end{center}
\end{figure}

In Fig. \ref{fig:pd2}, we show the phase diagram for the ripplopolaron system
on the helium surface in the variables $\left(  n_{0},\alpha\right)  $ for two
temperatures $T=0.1%
\operatorname{K}%
$ and $T=1%
\operatorname{K}%
$, keeping other parameters the same as described above. At the lower
temperature, we can see in Fig. \ref{fig:pd2} (\emph{a}) three regimes for the
ripplonic polaron system: the polaron Wigner solid, the electron Wigner solid
and the electron liquid. At low temperatures, these three regimes consequently
follow each other when increasing the electron concentration. The critical
concentration for the transition between the polaron and electron Wigner
solids monotonically increases with an increasing coupling strength. The other
critical concentration, which indicates melting of the electron Wigner solid
into the electron liquid exhibits a non-monotonic behavior as a function of
$\alpha$. At small $\alpha$, electron-ripplon scattering favors the melting of
the Wigner crystal, which can be explained by electron-ripplon scattering. On
the contrary, for larger $\alpha$ the electron-ripplon interaction favors the
Wigner crystallization because of increase in effective mass of the polarons.%

At the higher temperature $T=1%
\operatorname{K}%
$, close to the experimental conditions of Ref. \cite{Rees-PRL-2011}, all four
phases of the ripplopolaron system can be observed in the range of densities
$3\times10^{7}%
\operatorname{cm}%
^{-2}\lessapprox n_{0}\lessapprox3\times10^{10}%
\operatorname{cm}%
^{-2}$, as seen from Fig. \ref{fig:pd2} (\emph{b}). When increasing the
coupling strength, the electron liquid can turn into a polaron liquid, and the
electron Wigner solid can transform to the polaron Wigner solid, as expected.
Remarkably, at the relatively high temperature $T=1%
\operatorname{K}%
$, both polaron and electron liquids crystallize, respectively, to polaron and
electron Wigner crystals when the electron concentration \emph{increases},
contrary to the low-temperature case. This change of sequence of phases
between ripplopolaron systems at lower and higher temperatures finds a
transparent physical explanation through the interplay of the following
factors. On one hand, at high temperatures the formation of a polaron dimple
can be favored by the strengthening of the confinement potential, because the
thermal fluctuations of the electron motion become gradually more restricted
by the neighboring electrons when decreasing the inter-electron distance. On
the other hand, at low temperatures, when thermal fluctuations are less
important, melting of the electron Wigner solid can be favored by
zero-temperature quantum fluctuations of the electron motion: this is the case
of \emph{quantum melting}. We do see quantum melting at nonzero temperature,
and it is expected to persist down to $T=0$. This explains the different
sequence of phases between low-temperature and high-temperature regimes for
the Wigner solid.

{\normalsize At very high densities, the Fermi energy of electrons can be
comparable with their averaged kinetic energy and, consequently, quantum
melting of the Wigner crystal can be strongly influenced by the Fermi
statistics. Using the material parameters described above, this range of
densities is estimated as }$n_{0}\gtrsim10^{11}%
\operatorname{cm}%
^{-2}${\normalsize . In Ref. \cite{Gunzler}, quantum melting of an electron WS
to a degenerate Fermi gas was experimentally detected at }$n_{0}\sim10^{11}%
\operatorname{cm}%
^{-2}${\normalsize , confirming our estimations. We do not consider here
the electron-ripplon system at very high densities, when quantum melting
occurs to a degenerate Fermi gas. This regime will be a subject of the further
study.}

\begin{figure}
[th]
\begin{center}
\includegraphics[
height=5.9577in,
width=3.0943in
]%
{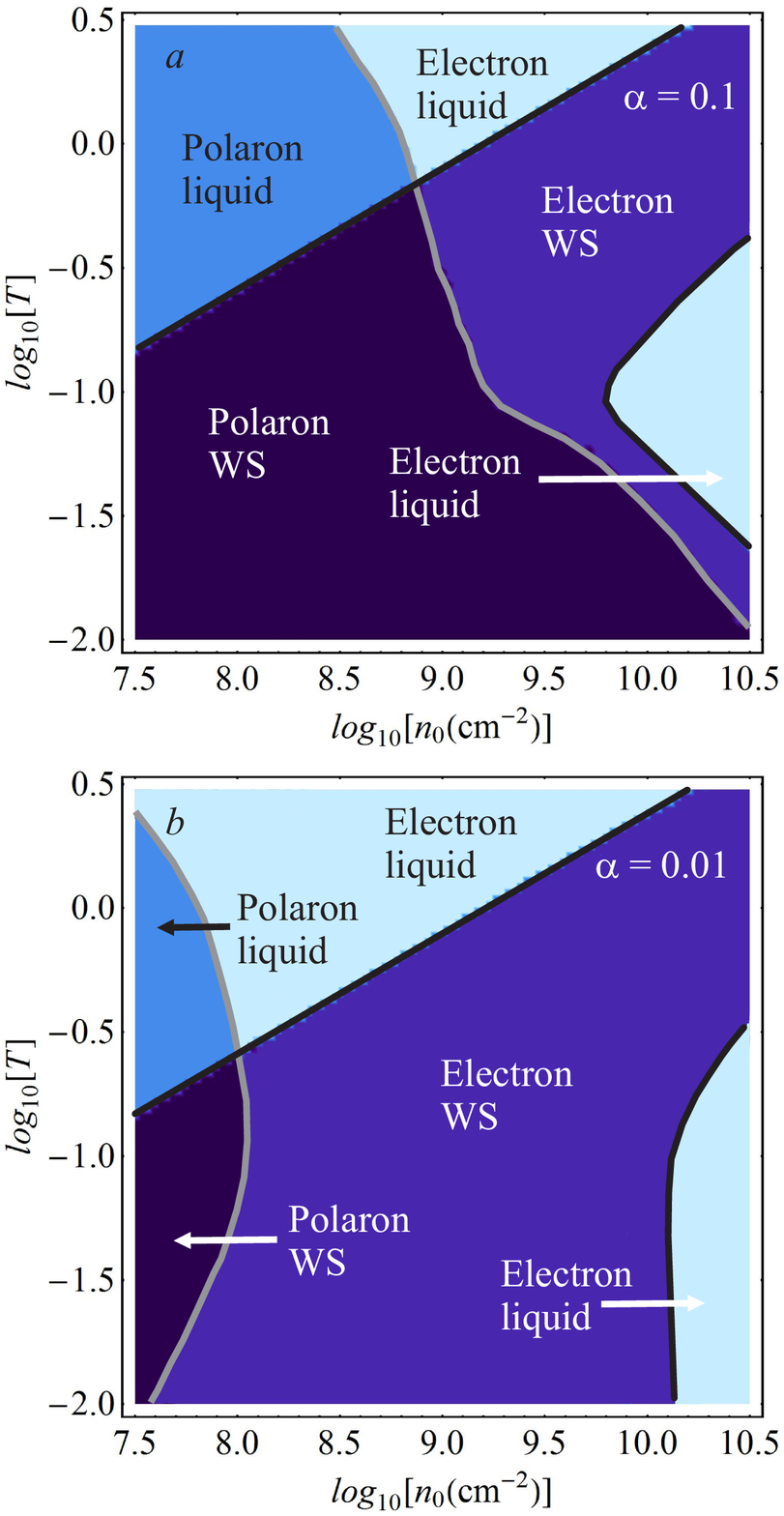}%
\caption{Phase diagram for the ripplopolaron system on the helium surface in
the variables $\left(  n_{0},T\right)  $ for the coupling strength
$\alpha=0.1$ (\emph{a}) and $\alpha=0.01$ (\emph{b}) , with the helium film
width $h=1\operatorname{\mu m}$.}%
\label{fig:pd3}%
\end{center}
\end{figure}

Finally, Fig. \ref{fig:pd3} shows the phase diagram for the ripplopolaron
system on the helium surface in the variables $\left(  n_{0},T\right)  $
plotted using two values of the electron-ripplon coupling constant
$\alpha=0.1$ and $\alpha=0.01$. In analogy with the phase diagrams plotted in
Fig. \ref{fig:pd2}, the sequences of different phases when varying the
electron concentration and temperature can be described and physically
explained in the following way. At small concentrations and low temperatures,
the system naturally turns into a polaron Wigner crystal. When increasing the
temperature while keeping the concentration constant, the polaron Wigner
crystal can either melt to a polaron liquid at low densities, or shed the
dimple and change to an electron Wigner crystal at higher densities. In the
former case, the breakdown of the polaron Wigner crystal occurs through the
melting of the lattice, but polaron dimples survive. In the latter case, the
polaron Wigner solid is changed to the electron Wigner solid through the
polaron dissociation. When temperature rises further, both the electron Wigner
crystal and the polaron liquid can change to the electron liquid but in a
different way: the electron Wigner crystal melts, while the polaron liquid
dissociates. When increasing the electron concentration at fixed temperature,
also the electron Wigner crystal can melt.%

Remarkably, the electron liquid phase appears not only for high temperatures
but also as an \textquotedblleft island\textquotedblright\ for low
temperatures and high electron concentrations (see Figs.~\ref{fig:pd2}%
(\emph{a}) and \ref{fig:pd3}). This means that the system displays a reentrant
melting transition from the electron liquid phase to the electron solid phase
at some fixed high $n_{0}$ when increasing the temperature or decreasing the
coupling strength. In other words, the system displays \textit{solidification
by heating} (for high enough electron concentrations). This sort of
transition, known as \textquotedblleft freezing by heating\textquotedblright%
\ transition, has been predicted for mesoscopic systems~\cite{vicsek,stanley}
and recently demonstrated for colloids driven by a non-uniform
force~\cite{tkachenko}. This counter-intuitive behavior does not violate
principles of thermodynamics and has been observed both in non-equilibrium and
equilibrium systems, see, e.~g., \cite{Greer,Lee2015}.

In Refs. \cite{vicsek,stanley,tkachenko}, the mechanism of the inverse melting
was explained by the fluctuation-driven increase of the effective size of the
particles (i.e., the area effectively occupied by the particle during its
fluctuation-driven random motion) in the molten state such that they form a
solid state with increasing temperature. For ripplonic polarons, the revealed
sequence of the reentrant electron phases when increasing temperature at high
electron concentrations can be explained in other way: by the fact that the
melting phase transition is differently driven by quantum and thermal
fluctuations. In (\ref{Rc}) and (\ref{rho}), \emph{thermal} fluctuations
contribute to the temperature dependence of $\left\langle \mathbf{R}_{c}%
^{2}\right\rangle $ and $\left\langle \boldsymbol{\rho}^{2}\right\rangle $
through the distribution functions $\coth\left(  \beta\hbar\Omega
_{j}/2\right)  $. The contribution of \emph{quantum} fluctuations to the
temperature dependence of the averaged squared radii occurs through the
polaron parameters $\left\{  \Omega_{j},w\right\}  $ (which also depend on
temperature). The electron-ripplon interaction can become effectively stronger
with rising temperature in some range of temperatures, and hence the
ripplon-induced potential for an electron becomes deeper and narrower in that
range. Quantum fluctuations may then favor to a non-monotonic dependence of
$\left\langle \mathbf{R}_{c}^{2}\right\rangle $ as a function of temperature.
On the contrary, thermal fluctuations always contribute to increase
$\left\langle \mathbf{R}_{c}^{2}\right\rangle $. Thus the transition
temperature can result from an interplay of quantum and thermal fluctuations.

When increasing the temperature, the area of the reentrant melting transition
shifts to higher concentrations, so that is not seen in Fig. \ref{fig:pd2}%
(\emph{b}) but appears in Fig. ~\ref{fig:pd2}(\emph{a}). Comparing the phase
diagrams for two coupling strengths, we can note that the region of stability
for the electron Wigner crystal substantially expands with decreasing $\alpha
$.

{\normalsize As mentioned above, electron-ripplon coupling in the experimental
conditions of Refs. \cite{Rees-PRL-2011,Rees-PRL-2012,Rees-JPSJ} (where a
stable Wigner crystal has been detected) is rather weak, }$\alpha\sim
1.5\times10^{3}${\normalsize . Hence the obtained phase diagrams are in
agreement with these experiments. We can also suggest that the Wigner solid in
the experiments \cite{Rees-PRL-2011,Rees-PRL-2012,Rees-JPSJ} can be classified
as an electron Wigner solid. For the experiments where the melting point of
the electron Wigner lattice at the helium surface was determined using
measurements of the mobility and the microwave response \cite{Jiang,Mistura},
the thickness of the helium film was significantly smaller than in the phase diagrams
calculated in the present work. However these experiments are also related to
the very weak-coupling polaron regime, where the electron WS rather than the
polaron WS can exist.}

{\normalsize It should be noted that the film thickness at high densities can
be strongly reduced \cite{Etz,Hu}. Consequently, at electron densities of
}$10^{10}${\normalsize  cm}$^{-2}$ {\normalsize and higher the helium film can
hardly have a thickness }$h=1%
\operatorname{\mu m}%
${\normalsize . Therefore some portions of the phase diagrams shown in Figs.
\ref{fig:pd1} to \ref{fig:pd3} are experimentally not accessible. However they
represent a theoretical interest for a many-polaron problem.}

\section{Conclusions}

In the present work, we have analyzed different phases of a ripplonic polaron
system on the surface of a liquid helium film, and the behavior of these
phases when varying parameters of the system: the temperature, the electron
concentration, and the electron-ripplon coupling strength. The
electron-ripplon system is considered in a wide range of electron densities
leaving out very high densities where the Fermi statistics becomes important.
The treatment has been performed within the arbitrary-coupling and
finite-temperature variational path-integral formalism based on the
Jensen-Feynman variational principle for the free energy.

We demonstrated that, by varying the electron-ripplon coupling strength
$\alpha$ and other parameters such as the electron concentration $n_{0}$ on
the helium surface and temperature $T$, in the vicinity of typical
experimental values, the system displays a rich phase behavior. We have found
that the experimental conditions corresponding to Ref. \cite{Rees-PRL-2011}
are favorable for the electron Wigner solid decoupled from the dimple lattice
rather than for other phases. This conclusion is in agreement with the
observation of an electron Wigner solid in that work.

For a set of typical experimental
parameters~\cite{Rees-PRL-2011,Rees-PRL-2012,Rees-JPSJ} $h=1%
\operatorname{\mu m}%
$ and $n_{0}=2.58\times10^{9}%
\operatorname{cm}%
^{-2}$, we revealed four different phases: (i) the electron solid (lattice),
at low temperatures and weak couplings; (ii) the polaron Wigner solid, at low
temperatures and strong couplings; (iii) the electron liquid, at high
temperatures and weak couplings; and (iv) the polaron liquid phase, when the
polaron Wigner solid melts but the electrons do not decouple from the dimples,
due to the strong electron-ripplon coupling. {\normalsize Remarkably, it
should be possible to observe all the predicted phases} at typical
experimental temperatures close to $T=1$%
~K~\cite{Rees-PRL-2011,Rees-PRL-2012,Rees-JPSJ}, for varying electron
concentration and the electron-ripplon coupling strength. For lower
temperatures, some of the phases disappear, like the polaron liquid phase.

Thus in addition to three phases of an electron-ripplon system which were
studied in the literature, the phase of a \emph{ripplopolaron liquid} is
possible at a combination of sufficiently low electron densities, strong
electron-ripplon couplings and high temperatures $T\sim1%
\operatorname{K}%
$. This regime is accessible experimentally, because all these parameters can
be controlled, including the coupling strength. Therefore we can expect for an
experimental detection of a ripplopolaron liquid.

The system displays even more striking phase behavior when varying the surface
electron concentration $n_{0}$ and temperature $T$. For weak electron-ripplon
couplings $\alpha$, the electron liquid phase dominates over a broad range of
$n_{0}$ and $T$, although the other above phases can also be observed, like
the polaron solid and liquid phases at low electron concentrations. The
polaron or electron liquid phase appears at rather high temperatures and can
crystallize with increasing $n_{0}$ (turning to, respectively, the polaron or
electron electron solid). For high electron concentrations, the system
exhibits quantum melting accompanied by an unusual reentrant behavior, i.e.,
the transition from a liquid to solid electron state with \textit{increasing}
temperature. This transition is known as the \textquotedblleft freezing by
heating\textquotedblright\ transition when fluctuations result in a
solidification of a molten state.

{\normalsize The mean-field Wigner approximation used in the present work was
successfully applied to polaron Wigner lattices in ionic crystals
\cite{FratiniEPJB14,Rastelli}. The application of this mean-field approach to
the WS on a liquid helium surface needs however some care. There exists an
enhancement of coupling constant in the Wigner solid phase due to the Bragg
scattering of ripplons \cite{fisher1979,Namaizawa,Monarkha1}. The Bragg
scattering of ripplons from the electron lattice gives rise to an enhanced
deformation of the dimple lattice \cite{ReesA}. This effect can be taken into
account, e.g., by introducing an effective Debye-Waller factor in the polaron
action (\ref{S}). The effect of the Bragg scattering can be especially
important at rather weak couplings. It can lead to a quantitative change of
the polaron energy and, consequently, to some shift of the boundaries on the phase diagrams. However, it can hardly change the physical picture of phases of the
rippopolaron system.}

Therefore, we demonstrated the existence of different phases of electrons and
polarons on surface of liquid helium, and we analyzed the regions of their
stability and the transitions between the revealed phases. Our findings
provide a deeper understanding of the phase behavior of the Wigner matter and
could be useful for the interpretation of the experimental observations such
as the temperature behavior of the decoupling transition (the decoupling of
the Wigner solid from the dimples). Furthermore, we expect that the revealed
unusual phases (like polaron liquid) and phase transitions (like the reentrant
electron lattice melting) could stimulate further studies, including new experiments.

\section{Acknowledgements}

We thank A. S. Mishchenko and D. G. Rees for valuable discussions. This research has been supported by the Flemish Research Foundation (FWO-Vl), project nrs. G.0115.12N, G.0119.12N, G.0122.12N, G.0429.15N, by the Scientific
Research Network of the Research Foundation-Flanders, WO.033.09N, and by the Research Fund of the University of Antwerp.

\end{document}